# Symmetry-Driven Electrical Switching of Anisotropic Skyrmion Hall Effect in Altermagnets

Wenhui Du, Kaiying Dou, Ying Dai*, Zeyan Wang, Baibiao Huang, Yandong Ma*

School of Physics, State Key Laboratory of Crystal Materials, Shandong University, Shandanan Street 27, Jinan 250100, China

*Corresponding author: daiy60@sina.com (Y.D.); yandong.ma@sdu.edu.cn (Y.M.)

**Abstract**

Controlling the skyrmion Hall effect (SkHE) is pivotal for developing topological spintronics but typically relies on magnetic field reversal. Here, we demonstrate a general strategy for the purely electrical switching of the SkHE in two-dimensional altermagnets. Through symmetry and model analysis, we reveal that the intrinsic altermagnetic symmetry imposes sublattice-dependent anisotropic exchange and Dzyaloshinskii-Moriya interactions. These interactions induce a highly anisotropic SkHE, where the transverse velocity is strictly dictated by the current direction relative to the crystal axes. Crucially, we show that an external electric field can strongly modulate these interaction parameters by inversing the altermagnetic symmetry, allowing for the reversible inversion of the anisotropic SkHE. Using first-principles and atomistic spin model simulations, this mechanism is further demonstrated in monolayer CaMnSn. Our study establishes a unique strategy for realizing precise, electrically tunable skyrmion transport without magnetic fields.

**Introduction**

Magnetic skyrmions—topologically protected spin textures characterized by a quantized winding number Q—have emerged as compelling candidates for next-generation information carriers [1-5]. Their localized particle-like nature, combined with robust topological stability, offers a fertile ground for exploring emergent electrodynamics in condensed matter [6-10]. Recently, the experimental realization of long-range magnetic order in two-dimensional (2D) van der Waals lattices [11-13] has reinvigorated this field, providing an ideal platform to engineer skyrmions with unprecedented tunability and density [14-30]. In this 2D limit, understanding and manipulating the current-driven dynamics of skyrmions becomes paramount for translating their exotic topological properties into functional spintronic logic and memory devices.

A hallmark of this dynamics is the skyrmion Hall effect (SkHE), where the topology-dependent Magnus force drives the spin texture transversely to the current direction [31,32]. While harnessing this transverse motion is crucial for topological spintronics, establishing a mechanism to reversibly switch this effect remains a formidable hurdle. In conventional ferromagnets (FMs), the SkHE is intrinsically tied to the fixed topological charge and is predominantly switched via external magnetic fields [32,33]—an approach inherently plagued by high power consumption and severe bottlenecks in local device integration [34]. Conversely, in antiferromagnetic (AFM) systems, the topological charge cancellation suppresses the Magnus force entirely, constraining skyrmions to linear trajectories and effectively extinguishing the SkHE [35,36]. Therefore, realizing a mechanism that permits efficient, purely electrical control of the SkHE—thereby breaking this FM-AFM dichotomy—stands as a critical imperative for the advancement of topological spintronics.

In this letter, we address this challenge by establishing a general mechanism for the reversible electrical switching of the SkHE in 2D altermagnets. Through symmetry and model analysis, we identify that the intrinsic crystal symmetry imposes sublattice-dependent anisotropic exchange and Dzyaloshinskii-Moriya interactions (DMI). These interactions give rise to a highly anisotropic SkHE, where the transverse velocity is strictly dictated by the current direction relative to the crystal axes. Crucially, we demonstrate that an external electric field can significantly modulate these anisotropic parameters through coupling to the altermagnetic symmetry, triggering a sign reversal of SkHE. Furthermore, we also validate this non-magnetic switching mechanism in monolayer CaMnSn via first-principles calculations and atomistic spin model simulations. Our findings not only demonstrate the feasibility of electrical control but also establish altermagnets as a promising platform for topological spintronics.

## Results and Discussion

We begin by clarifying the symmetry constraints on skyrmion properties in conventional 2D antiferromagnets. In these systems, the two FM sublattices (denoted as A and B) are typically related by combined time-reversal and spatial symmetries (e.g., $[C_2\|P]$ and $[C_2\|t]$), which forces the constituent skyrmions to carry opposite topological charges ($Q_A = -Q_B$) [20]. The current-driven motion is governed by the Thiele equation $\mathbf{G}\times(\mathbf{v}-\mathbf{u})+\mathcal{D}(\alpha\mathbf{v}-\beta\mathbf{u})+\nabla V(\mathbf{r})=0$ [35], where $G=(0,0,g)$ is the gyrovector with $g=-4\pi Q$. Here, the topological charge $Q$ is described as $Q=\frac{1}{4\pi}\int \mathbf{m}\cdot\left(\frac{\partial \mathbf{m}}{\partial x}\times\frac{\partial \mathbf{m}}{\partial y}\right)dxdy$ [6,37]. $\mathbf{v}$ denotes the drift velocity, while $\mathbf{u}=(u_x,u_x)=(u\cos\theta, u\sin\theta)$ represents the current-induced velocity with $\theta$ defining current angle relative to the $x$-axis. $\mathcal{D}$ characterizes the dissipative tensor, with $\alpha$ and $\beta$ being the Gilbert damping constant and non-adiabatic torque strength, respectively. $V(\mathbf{r})$ is the confining potential arising from the sample boundaries. Solving for the velocity $\mathbf{v}$, the general solution depends explicitly on the dissipative tensor components:

$$\mathbf{v}=\begin{cases}v_x\\v_y\end{cases}=\begin{cases}V_0[(g^2+\alpha\beta\mathcal{D}_{xx}\mathcal{D}_{yy})\cos\theta+(\beta-\alpha)g\mathcal{D}_{yy}\sin\theta]\\V_0[(\alpha-\beta)g\mathcal{D}_{xx}\cos\theta+(g^2+\alpha\beta\mathcal{D}_{xx}\mathcal{D}_{yy})\sin\theta]\end{cases} \quad (1)$$

where $V_0=u/(\alpha^2\mathcal{D}_{xx}\mathcal{D}_{yy}+g^2)$.

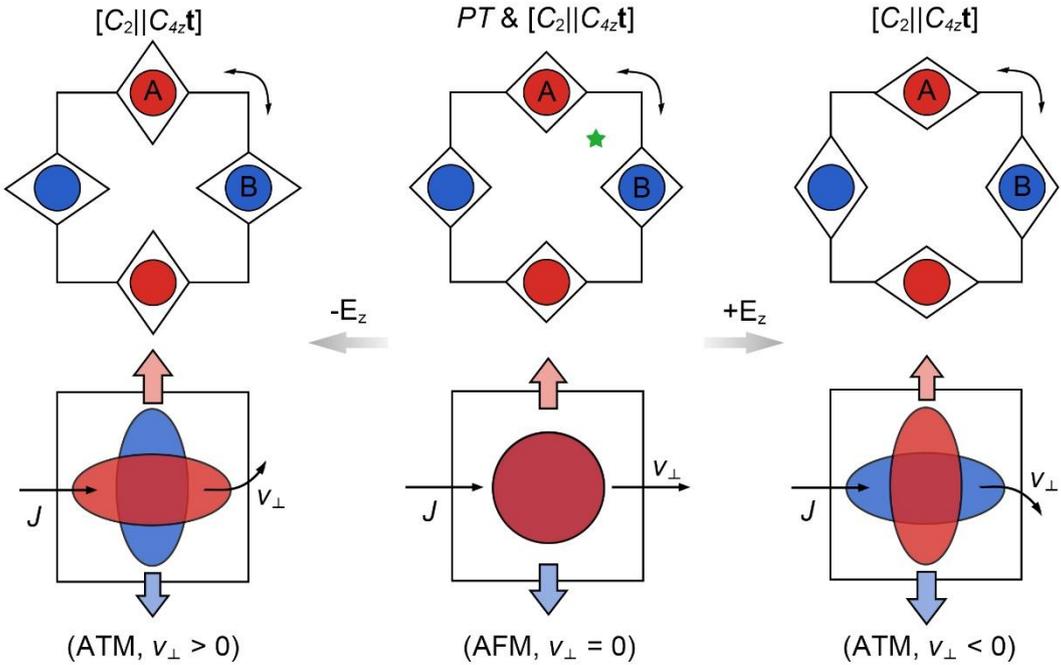

**Figure 1** Symmetry-driven electrical switching of the anisotropic SkHE. (Center) The pristine AFM
3

state is protected by [$C_2\|P$] or [$C_2\|t$] symmetry (green star), where the isotropic magnetic environment enforces an exact cancellation of transverse forces, confining skyrmions to linear trajectories ($v_\perp = 0$). (Left, Right) Application of an external electric field ($\mp E_z$) breaks the inversion symmetry, driving the system into ATM states. This induces sublattice-dependent anisotropy (depicted by the orthogonal red and blue ellipses), generating a finite, reversible transverse velocity ($v_\perp \neq 0$). The red and blue arrows indicate the transverse driving forces acting on sublattices A and B, respectively. Note that reversing the $E_z$-field direction inverts the transport anisotropy and the resulting anisotropic SkHE.

Crucially, the [$C_2\|P$] or [$C_2\|t$] symmetry enforces an isotropic magnetic environment (**Figure 1**, middle panel), strictly constraining the dissipative tensor for each sublattice to be diagonal and uniform ($\mathcal{D}_{xx} = \mathcal{D}_{yy} = \mathcal{D}$). Consequently, the transverse velocity for a single sublattice simplifies to $v_\perp = ug\mathcal{D}(\beta - \alpha)/(\alpha^2\mathcal{D}^2 + g^2)$ (detailed in **Note 1** of Supporting Information). However, since $g_A = -g_B$, the transverse velocities of the coupled sublattices cancel exactly ($v_\perp^A = -v_\perp^B$). This symmetry-protected cancellation confines skyrmions to linear trajectories along the current direction, establishing a fundamental barrier to transverse control that necessitates symmetry breaking.

To dismantle this symmetry barrier, the strict cancellation of the transverse velocities between sublattices must be lifted. We propose that an out-of-plane electric field ($E_z$) serves as the ideal symmetry breaker. It eliminates the global [$C_2\|P$] or [$C_2\|t$] symmetry but retains a joint rotational symmetry (e.g., [$C_2\|C_{4z}t$]), effectively transforming the isotropic AFM into an electric-field-induced ATM state. In this sense, the isotropic interactions on each sublattice are transformed into anisotropic ones.

We here consider a model 2D tetragonal lattice with $D_{4h}$ symmetry. The application of $E_z$ induces a coordination inequivalence between sublattices A and B (**Figure 1**, side panels), rendering the magnetic texture on each sublattice anisotropic (elliptical) rather than isotropic. Crucially, due to the 90° relative rotation connecting the sublattices, their dissipative tensors satisfy an orthogonal relationship: $\mathcal{D}_{xx}^A = \mathcal{D}_{yy}^B = \mathcal{D}_1$ and $\mathcal{D}_{yy}^A = \mathcal{D}_{xx}^B = \mathcal{D}_2$ [38-40]. Solving the Thiele equation under these anisotropic constraints yields the longitudinal ($v_\parallel$) and transverse ($v_\perp$) velocities of the composite skyrmion:

$$v_\parallel = V_0[2(g^2 + \alpha\beta\mathcal{D}_1\mathcal{D}_2) + g(\alpha - \beta)(\mathcal{D}_1 - \mathcal{D}_2)\sin 2\theta] \quad (2)$$

$$v_\perp = -V_0[g(\alpha - \beta)(\mathcal{D}_1 - \mathcal{D}_2)\cos 2\theta] \quad (3)$$

These expressions clearly indicate that an anisotropic transverse velocity is instantaneously generated (i.e., realizing the anisotropic SkHE), driven explicitly by the anisotropy difference ($\mathcal{D}_1 - \mathcal{D}_2$).

The hallmark of this mechanism is its electrical reversibility. Reversing the electric field direction ($E_z$ to $-E_z$) interchanges the local crystal environments of the two sublattices (**Figure 1**, side panels). This operation effectively swaps the dissipative components ($\mathcal{D}_1 \leftrightarrow \mathcal{D}_2$), which imposes a sign reversal on the anisotropy term $[(\mathcal{D}_1 - \mathcal{D}_2) \to (\mathcal{D}_2 - \mathcal{D}_1)]$. Consequently, the transverse velocity undergoes an immediate inversion:

$$v_\perp(-E_z) = -v_\perp(+E_z) \quad (4)$$

This result rigorously demonstrates that the anisotropic SkHE in such field-induced altermagnets is not only non-zero but also electrically switchable.

To validate this symmetry-derived mechanism in a realistic material system, we perform first-principles calculations and atomistic spin model simulations on monolayer CaMnSn [41]. As illustrated in **Figure 2(a)**, the system crystallizes in a tetragonal lattice (space group *P4/nmm*, point group *D$_{4h}$*), consisting of five atomic layers stacked in a Ca-Sn-Mn-Sn-Ca sequence. The unit cell contains two Mn atoms (Mn$_A$ and Mn$_B$) residing within the same atomic plane, which belong to magnetic sublattices A and B. These two sublattices are related by the spatial inversion symmetry (***P***), which ensures that the two nearest-neighbor Mn atoms experience equivalent and isotropic coordination environments. The optimized lattice parameter is 4.47 Å. Furthermore, the dynamic stability of the monolayer is confirmed by the absence of imaginary frequencies in the phonon spectrum (see **Figure S2** in Supplemental Material).

The macroscopic magnetic properties of this monolayer are strictly governed by the local electronic environment of Mn atoms. Formally, charge transfer establishes a nominal Mn$^{2+}$ oxidation state with $3d^5 4s^0$ configuration. Subject to the local tetrahedral crystal field exerted by the neighboring Sn atoms, the Mn-*3d* orbitals split into two manifolds: the lower-energy doublet *e$_g$* orbitals and the higher-energy triplet *t$_{2g}$* orbitals. Driven by strong intra-atomic Hund's exchange, these five valence electrons adopt a high-spin configuration, theoretically yielding a magnetic moment of 5 $\mu_B$ per Mn atom. Consistent with this localized physical picture, our first-principles calculations reveal a total magnetic moment of 8.95 μ$_B$ per unit cell. This value aligns well with the expected high-spin state, where the slight reduction from the ideal 10 μ$_B$ (for two Mn atoms) reflects the characteristic *p-d* hybridization between the Mn and adjacent Sn atoms, leaving the magnetization predominantly localized at the coplanar Mn sites.

To determine the magnetic ground state, we compare the total energies of spin-parallel and spin-antiparallel configurations, finding that AFM coupling is energetically favorable. As shown in **Figure S3(a)**, the band structure of monolayer CaMnSn exhibits complete spin degeneracy due to the preserved [$C_2 \| P$] symmetry relating the two sublattices. Application of an out-of-plane electric field ($E_z$) explicitly breaks this symmetry. While the zero net magnetization is strictly preserved, $E_z$ lifts the sublattice degeneracy, inducing a pronounced spin-polarized band splitting that is the hallmark of altermagnetism [42-44] [see **Figure 2(d)**]. Specifically, the bands along M-Y-Γ and Γ-X-M paths remain degenerate in energy but are completely segregated into opposite spin channels. This distinctive behavior stems from the breaking of [$C_2 \| P$] symmetry while preserving diagonal mirror symmetry ($M_Φ$), which enforces spin-momentum locking [42-44]. Remarkably, reversing the electric field direction (−$E_z$) interchanges the local environments of the sublattices, effectively inverting the sign of the momentum-dependent spin splitting [**Figure S3(b)**].

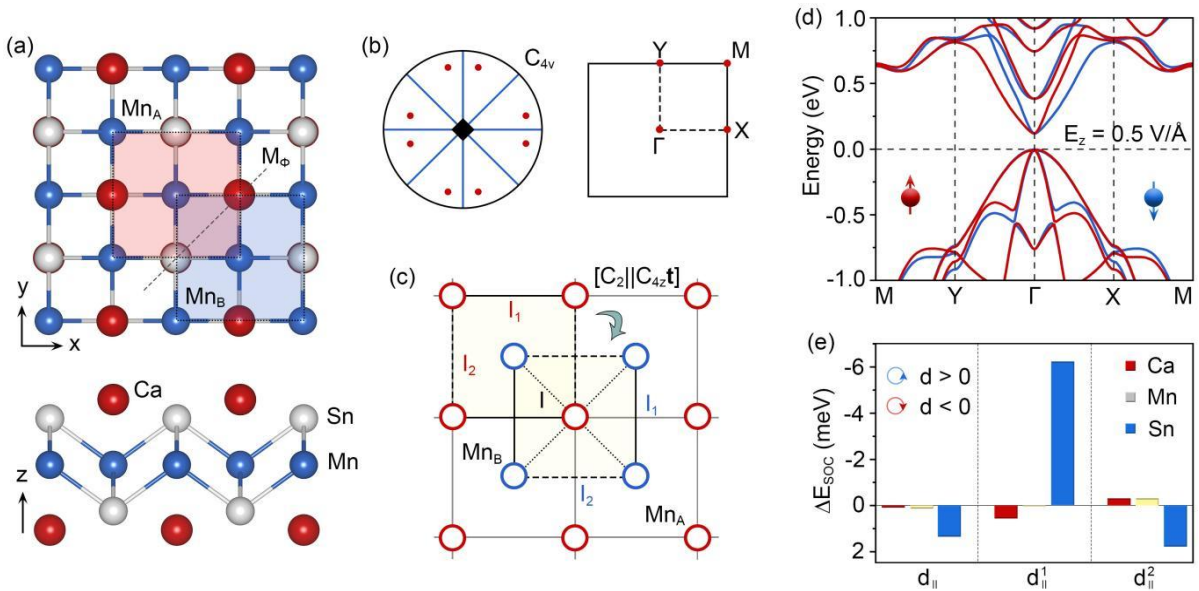

**Figure 2** Crystal structure, symmetry, and field-induced altermagnetic properties of monolayer CaMnSn. (a) Crystal structure of monolayer CaMnSn from top and side views. (b) Equatorial plane

projection of the point group for monolayer CaMnSn under an out-of-plane electric field (left panel) and the corresponding 2D Brillouin zone with high-symmetry paths (right panel). (c) Schematic of the magnetic interactions ($I$ represents $J$ or $D$) in the field-induced altermagnetic state. The interactions $I$, $I_1$, and $I_2$ are indicated by dotted, solid, and dashed lines, respectively. (d) Spin-resolved band structure of monolayer CaMnSn under $E_z = 0.5$ V/ Å. (e) SOC energy projected on atoms ($\Delta E_{SOC}$) for monolayer CaMnSn under $E_z = 0.6$ V/ Å.

To investigate the magnetic interactions in monolayer CaMnSn, we construct the following Heisenberg spin Hamiltonian:

$$H = -\sum_{<i,j>} J\mathbf{S}_i \cdot \mathbf{S}_j - \sum_n \sum_{<i,j>_n} J_n \mathbf{S}_i \cdot \mathbf{S}_j - \sum_{<i,j>} \mathbf{D} \cdot (\mathbf{S}_i \times \mathbf{S}_j)$$
$$- \sum_n \sum_{<i,j>_n} \mathbf{D}_n \cdot (\mathbf{S}_i \times \mathbf{S}_j) - K \sum_i (S_i^z)^2 \qquad (5)$$

Here, $\mathbf{S}_i$ indicates the unit vector representing the local spin at the $i$-th Mn site. $\langle i,j \rangle$ and $\langle i,j \rangle_n$ run over all inter-sublattice Mn atomic pairs and intra-sublattice Mn atomic pairs, respectively. $J$ and $J_n$ ($n$ = 1, 2) denote the Heisenberg exchange interactions between the inter-sublattice and intra-sublattice sites, respectively. $\mathbf{D}$ and $\mathbf{D}_n$ ($n$ = 1, 2) represent the inter-sublattice and intra-sublattice DMI, respectively. The index $n$ is introduced to differentiate the sublattice-specific intra-sublattice interactions stemming from the coordination inequivalence between the two sublattices, as illustrated in **Figure 2(c)**. The last term is the single-ion anisotropy (SIA).

As illustrated in **Figure S4**, we consider four different magnetic configurations to obtain the exchange interactions in monolayer CaMnSn. The derived exchange parameters are $J = -47.55$ meV and $J_1 = J_2 = 7.34$ meV, respectively. These values indicate that the inter-sublattice exchange interaction favors AFM coupling, while FM coupling dominates the intra-sublattice interactions. Regarding the DMI, the $D_{4h}$ (*P4/nmm*) symmetry of tetragonal lattice dictates that $\mathbf{D}_{ij}$ vector between neighboring intra-sublattice Mn atoms is perpendicular to their connecting bond [45]. It can be expressed as $\mathbf{D}_{ij} = d_\parallel (\mathbf{z} \times \mathbf{u}_{ij}) + d_\perp \mathbf{z}$, where $\mathbf{z}$ is the unit vector along the *z*-axis and $\mathbf{u}_{ij}$ is the unit vector pointing from site *i* to site *j*; see **Figure S5**. Since the Ca-Mn-Sn bond lies within a mirror plane, the out-of-plane component $d_\perp$ can be neglected, leaving only the in-plane component $d_\parallel$ to be considered. Moreover, the crystal symmetry further mandates that the in-plane components $d_\parallel^1$ and $d_\parallel^2$ possess opposite signs. Our calculations yield $d_\parallel^1 = -d_\parallel^2 = -0.42$ meV, where the negative sign corresponds to clockwise chirality. The inter-sublattice DMI is found to be zero, which is forbidden by the space-inversion symmetry. Regarding the SIA, the calculated parameter $K$ is -0.11 meV, indicating an in-plane magnetic anisotropy for monolayer CaMnSn.

We further investigate the evolution of these magnetic interactions under an electric field $E_z$ ranging from 0.1 to 0.6 V/Å. The application of $E_z$ explicitly breaks the inversion symmetry, reducing the crystal symmetry from $D_{4h}$ to $C_{4v}$. As detailed in **Table S1**, the inter-sublattice exchange interaction $J$ exhibits negligible variation, maintaining a robust AFM coupling. In stark contrast, the degeneracy of the intra-sublattice exchange interactions is lifted; $J_1$ and $J_2$ become non-equivalent but remain positive, preserving the FM ordering for intra-sublattice coupling. Similarly, the intra-sublattice DMI components $d_\parallel^1$ and $d_\parallel^2$ split in magnitude while retaining opposite signs. Crucially, as illustrated in **Figure 2(c)**, sublattices A and B are no longer related by the $[C_2 \Vert P]$ symmetry operation, but are instead interconnected via $[C_2 \Vert C_{4z}t]$. This symmetry reduction naturally gives rise to sublattice-specific anisotropic interactions—a definitive hallmark of the electric-field-induced altermagnetic state.

A profound consequence of the electric-field-induced inversion-symmetry breaking is the emergence of a non-zero inter-sublattice DMI oriented perpendicular to the bond connecting two adjacent Mn atoms. We focus on the in-plane component $d_\parallel$, as the out-of-plane component does not contribute to the formation of non-coplanar spin textures. As summarized in **Table S1**, $d_\parallel$ increases monotonically with $E_z$. To elucidate its microscopic origin, we plot the atomically resolved spin-orbit coupling (SOC) energy difference $\Delta E_{\text{SOC}}$ [extracted from spin-spiral configurations, see **Figures S6** and **S7**] in **Figure 2(e)**. It can be seen that the DMI is not primarily contributed by the Mn atoms, but is predominantly driven by the strong SOC of the neighboring Sn atoms. This firmly confirms that the system follows a Fert–Levy type mechanism [46,47], where the heavy ligand mediates the DMI between magnetic sites. Concurrently, the single-ion anisotropy parameter $K$ decreases in magnitude with increasing $E_z$ (**Table S1**), indicating a gradual weakening of the in-plane magnetic confinement.

Upon reversing the electric field direction ($-E_z$), the system exhibits a mathematically elegant, symmetry-dictated response. While the inter-sublattice exchange $J$ and single-ion anisotropy $K$ remain invariant, the intra-sublattice terms undergo a complete spatial redistribution. Specifically, the values of $J_1$ and $J_2$ are precisely exchanged ($J_1 \leftrightarrow J_2$), as are the magnitudes of $d_\parallel^1$ and $d_\parallel^2$ (with their signs unchanged). Furthermore, the inter-sublattice DMI $d_\parallel$ undergoes a sign reversal while maintaining its magnitude, signifying a switch in the chirality, as illustrated in **Figure S5**. Consequently, reversing the electric field effectively swaps the strengths of the sublattice-specific anisotropic interactions. This computational result directly validates our phenomenological model ($\mathcal{D}_1 \leftrightarrow \mathcal{D}_2$), firmly cementing an electrically reversible pathway for controlling the anisotropic SkHE.

To investigate the spin textures of monolayer CaMnSn, atomistic spin model simulations are performed using the Landau-Lifshitz-Gilbert (LLG) equation [48,49], with the spin Hamiltonian parameterized by first-principles calculations. In the absence of an external electric field ($E_z = 0$), the system exhibits a conventional AFM order. The application of an out-of-plane electric field gradually suppresses the in-plane magnetic anisotropy, thereby promoting the emergence of non-coplanar spin textures. As illustrated in **Figure 3(a)**, isolated ATM bimerons are stabilized when $E_z$ reaches 0.5 V/Å. Upon further increasing the field to 0.6 V/Å, these bimerons undergo a topological transition and evolve into skyrmions [**Figure 3(b)**].

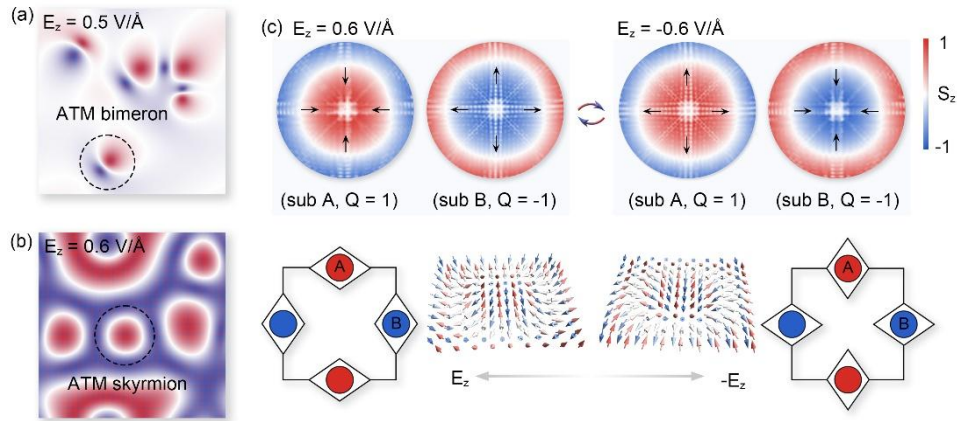

**Figure 3** Field-driven topological transitions and electrical control of altermagnetic skyrmions. Spin textures of monolayer CaMnSn under out-of-plane electric field $E_z$ of (a) 0.5 V/ Å and (b) 0.6 V/ Å. (c) Enlarged sublattice-resolved views of spin textures of monolayer CaMnSn under $E_z = \pm 0.6$ V/ Å. In (a-c), the color map encodes the out-of-plane spin components, while the arrows indicate the in-plane spin components. (d) Schematic illustrating the electrical control of altermagnetic skyrmions.

These non-coplanar ATM quasiparticles comprise two coupled FM sub-textures with opposite polarities. Taking the ATM skyrmion at $E_z = 0.6$ V/Å as an example, the FM skyrmion in sublattice A exhibits an upward polarity with a topological charge of $Q = 1$. whereas the skyrmion in sublattice B possesses a downward polarity with $Q = -1$, as shown in **Figure 3(c)**. Notably, despite their opposite topological charges, both sub-textures share an anticlockwise whirling chirality. This behavior is primarily dictated by the dominant intra-sublattice DMI $d_\parallel^1$ and inter-sublattice DMI $d_\parallel$. As illustrated in **Figure S5(a)**, while the intra-sublattice DMIs $d_\parallel^1$ and $d_\parallel^2$ in both sublattices favor anticlockwise and clockwise spin spiralings, respectively, the magnitude of $d_\parallel^1$ is markedly dominant. Combined with the inter-sublattice DMI $d_\parallel$, which further promotes an anticlockwise spin arrangement in two sublattices, the resulting chirality is uniquely determined. It is worth noting that while ATM skyrmions exhibit a structural resemblance to their AFM counterparts, their stability is originated from a distinct mechanism. Specifically, rather than relying on perfectly compensated isotropic interactions, ATM skyrmions are uniquely sustained by sublattice-specific anisotropic interactions, a direct consequence of the broken equivalence between the two magnetic sublattices in the ATM lattice.

Upon reversing the electric field to $-E_z$, the ATM skyrmion system exhibits a profound reconfiguration. As shown in **Figure 3(c)**, while the core polarities and topological charges ($Q = \pm 1$) of the FM sub-textures remain invariant, their spiral helicity undergoes a complete reversal from anticlockwise to clockwise. This helicity inversion is fundamentally rooted in the field-induced redistribution of the intra-sublattice DMI and the sign reversal of inter-sublattice DMI. Specifically, as illustrated in **Figure S5(b)**, the reversal of $E_z$ effectively swaps the magnitudes of $d_\parallel^1$ and $d_\parallel^2$ without changing their signs, thereby flipping the chirality of the dominant intra-sublattice DMI component. Concurrently, the sign of the inter-sublattice DMI $d_\parallel$ is reversed, leading to a direct reversal of its associated chirality. These two factors act in concert to drive the complete reversal of the spiral helicity across both sub-textures. More importantly, this field reversal simultaneously interchanges the localized anisotropic effective fields experienced by the skyrmions on the two sublattices. This precise spatial swapping of the anisotropic landscape firmly paves the way for the purely electrical control of skyrmion transport, including the deterministic switching of the SkHE.

To verify the electrical control of the SkHE in monolayer CaMnSn, we next investigate the current-driven dynamics of ATM skyrmions under $E_z = \pm 0.6$ V/Å. As illustrated in **Figure 4(a)**, injecting a current along the +x direction under $E_z = 0.6$ V/Å triggers a pronounced transverse deflection. The skyrmion trajectory deviates toward the -y direction, reaching a net transverse displacement of 1.6 nm over 800 ps. This observation of a non-zero skyrmion Hall angle originates directly from the uncompensated transverse velocities of the two magnetic sublattices. Driven by the sublattice-specific anisotropic fields, the FM sub-texture on sublattice B acquires a greater transverse velocity magnitude than that on sublattice A, culminating in a net macroscopic deflection along the -y axis, perfectly consistent with our phenomenological model (recall **Figure 1**).

Remarkably, reversing the applied electric field to $E_z = -0.6$ V/Å induces a complete reversal of the SkHE. As demonstrated in **Figure 4(a)**, under the same +x axis injection, the ATM skyrmion deflects toward the +y direction, manifesting a mirror-symmetric reflection of the trajectory across the longitudinal axis. The transverse displacement magnitude is perfectly conserved (reaching an identical absolute value at corresponding time steps), yet its spatial direction is exactly inverted. This deterministic switching of the SkHE is a direct dynamical consequence of the electric-field-induced swapping of the sublattice-specific anisotropic fields. To further validate this mechanism, we modeled a conventional AFM skyrmion by replacing the sublattice-specific anisotropic fields with isotropic interactions, which are preserved by the $[C_2 \| P]$ symmetry in the absence of an electric field. As shown in **Figure 4(a)**, when driven by a current along the +x direction, the skyrmion merely oscillates near its equilibrium without any transverse deflection. Together, these dynamical simulations unequivocally demonstrate that the field-induced, sublattice-resolved anisotropy is the fundamental physical origin for both the emergence and the precise electrical switching of the SkHE in two-dimensional altermagnets.

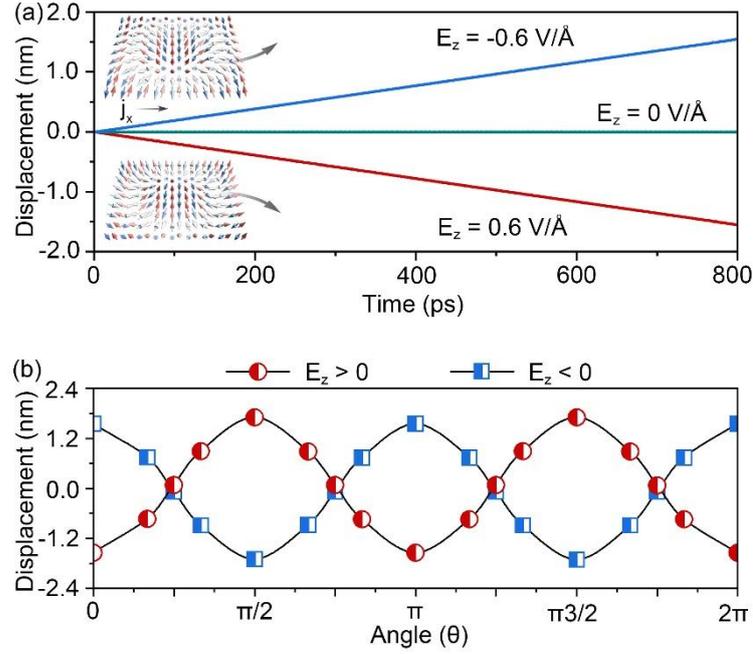

**Figure 4** Electrically switchable anisotropic SkHE. (a) Time evolution of the transverse displacement of skyrmion in monolayer CaMnSn without and with considering the effective sublattice-specific anisotropic field. (b) The relationship between ATM skyrmion transverse displacement and current injection angle θ under $E_z > 0$ and $E_z < 0$ states.

To systematically map this anisotropic transport, we evaluate the skyrmion transverse displacement as a function of the current injection angle ($\theta$) under $E_z = \pm 0.6$ V/Å. As shown in **Figure 4(b)**, the simulated angular dependence perfectly matches the analytical functional form predicted by Eqs. (3) and (4)—being proportional to $\cos 2\theta$—thereby corroborating the intrinsic anisotropic SkHE inherent to the field-induced ATM state. Remarkably, reversing the electric field direction systematically inverts the transverse velocity across all current directions. This ubiquitous sign reversal provides robust, omnidirectional electrical control over the SkHE trajectory, a feature highly promising for topological spintronics. Furthermore, **Figure 4(b)** shows that the SkHE strictly vanishes, with the transverse displacement dropping to zero, when the current injection angle satisfies $\theta = (2n+1)\pi/4$ ($n \in \mathbb{Z}$). At these specific symmetry-dictated orientations, the transverse velocity components of the two sublattices become exactly equal in magnitude but opposite in sign, leading to a perfect cancellation of the macroscopic Magnus-like force.

## Conclusion

In summary, we demonstrate a symmetry-driven mechanism for the reversible electrical switching of the SkHE in 2D altermagnets. An external electric field breaks the spatial equivalence of the magnetic sublattices, inducing anisotropic interactions that circumvent the strict topological charge cancellation of conventional antiferromagnets. Crucially, reversing the electric field deterministically swaps these sublattice-specific anisotropies, triggering a complete inversion of the transverse skyrmion trajectory. Validated in monolayer CaMnSn, this field-controlled dynamics bridges the gap between ferromagnetic and antiferromagnetic spintronics, offering a highly promising platform for energy-efficient, trajectory-encoded logic devices.

**Methods**

Our first-principles calculations are carried out using density functional theory (DFT) as implemented in the Vienna ab initio Simulation Package (VASP) [50,51]. The exchange-correlation interaction is treated within generalized gradient approximation (GGA) [52] using Perdew–Burke–Ernzerhof (PBE) functional. A plane-wave cutoff energy of 520 eV is employed, and the convergence criteria for energy and force are set to $10^{-6}$ eV and 0.01 eV/Å, respectively. The Monkhorst-Pack k-point mesh of $25 \times 25 \times 1$ is adopted to sample the 2D Brillouin zone. To account for strong correlation effects, following the previous works [41], the GGA + U approach is employed, with a Hubbard U value of 2 eV applied to the $3d$ electrons of the Mn atoms. The vacuum space of 20 Å along z-direction is included to eliminate interactions between periodic images. DFT-D3 method is applied to account for vdW interaction [53].

Atomistic spin model simulations are conducted using the VAMPIRE package [54] based on the spin Hamiltonian and Landau-Lifshitz-Gilbert (LLG) equation:

$$\frac{\partial S_i}{\partial t} = -\frac{\gamma}{(1+\lambda^2)}\left[S_i \times B_{eff}^i + \lambda S_i \times \left(S_i \times B_{eff}^i\right)\right] \qquad (6)$$

where $S_i$ is the unit vector representing the spin moment at site $i$, and $B_{eff}^i$ is the on-site effective magnetic field, given by $B_{eff}^i = -\frac{1}{u_s}\frac{\partial H}{\partial S_i}$, which governs the time evolution of spin structures. Here, $\gamma$ is the gyromagnetic ratio, and $\lambda$ is the damping constant. In the simulations, the system is gradually cooled from a high-temperature disordered state ($T \gg T_C$) to the investigated low temperature. Stable spin textures are obtained using a $500 \times 500 \times 1$ supercell with periodic conditions. The total simulation time is 3 ns, with a time step of 0.1 fs. The values of $\gamma$ and $\lambda$ are set to $1.76 \times 10^{11} T^{-1} s^{-1}$ and 1.0, respectively.

Spin dynamics simulations are performed using the MuMax3 package [55]. The simulation cell, with a size of $600 \times 600 \times 1$ nm³, is discretized using a mesh size of $1 \times 1 \times 1$ nm³. The parameters for anisotropic effective field are mapped from the magnetic parameters in 2D atomic spin

Hamiltonian model [56]. Notably, the inter-sublattice interactions are scaled down fivefold to accentuate the transverse displacement of ATM skyrmions for improved visibility. This adjustment serves to highlight the underlying physical mechanism without altering the qualitative nature of the SkHE.

## Acknowledgements

This work is supported by the National Natural Science Foundation of China (No. 12274261), Taishan Young Scholar Program of Shandong Province, and Shandong Provincial Natural Science Foundation of China (No. ZR2024QA016).

## Conflict of Interest

The authors declare no competing interests.